\documentclass{article}
\usepackage{graphicx}
\usepackage{epstopdf}
\usepackage{amsmath, amsthm, latexsym}
\usepackage{amssymb}
\usepackage{color}

\newtheorem*{remark}{Remark}
\usepackage{subfigure}

\usepackage{hyperref}

\usepackage[small, bf]{caption}

\begin{document}

\vspace*{3cm} \thispagestyle{empty}
\vspace{5mm}

\noindent \textbf{\Large{Embeddings and time evolution of the Schwarzschild wormhole}}\\

\noindent \textbf{\normalsize Peter Collas}\footnote{Department of Physics and Astronomy, California State University, Northridge, Northridge, CA 91330-8268. Email: peter.collas@csun.edu.}
\textbf{\normalsize and David Klein}\footnote{Department of Mathematics and Interdisciplinary Research Institute for the Sciences, California State University, Northridge, Northridge, CA 91330-8313. Email: david.klein@csun.edu.}\\

\vspace{4mm} \parbox{11cm}{\noindent{\small We show how to embed  spacelike slices of the Schwarzschild wormhole (or Einstein-Rosen bridge) in $\mathbb{R}^{3}$. Graphical images of embeddings are given, including depictions of the dynamics of this nontraversable wormhole at constant Kruskal times up to, and beyond, the ``pinching off'' at Kruskal times $\pm1$.}\vspace{5mm}

}\\
\vspace{6cm}
\pagebreak

\section[Intro]{Introduction}\label{Intro}

\noindent Schwarzschild's 1916 solution \cite{Schwarz} to the Einstein field equations is perhaps the most well-known of the exact solutions. In polar coordinates, the line element for a mass $m$ is,

\begin{equation}
ds^{2}=-\left(1-\frac{2m}{r}\right)dt^{2}+ \left(1-\frac{2m}{r}\right)^{-1}dr^{2}
+r^{2}(d\theta^{2}+\sin^{2}\theta d\phi^{2}),
\label{a1}
\end{equation}
where here and throughout we adopt units for which $G=1=c$. The realization that Eq. \eqref{a1} describes what is now called a black hole was far from immediate. It was not realized until the work of Oppenheimer and Snyder  \cite{oppen} in 1939 that such objects might actually exist, and that they could result from the collapse of sufficiently massive stars.\\

\noindent In contrast, investigations leading to the study of wormholes began almost immediately. Within one year of Einstein's final formulation of the field equations, Ludwig Flamm recognized that the Schwarzschild solution could represent what we now describe as a wormhole, and in the 1920s, Hermann Weyl speculated about related possibilities (cf. \cite{MT88}).\\  

\noindent Then in 1935 Einstein and Rosen \cite{rosen} introduced what would  be called the Einstein-Rosen bridge as a possible geometric model of particles that avoided the singularities of points with infinite mass or charge densities.  In the uncharged case, the bridge arises from the coordinate change, $y^{2}=r-2m$, which transforms Eq. \eqref{a1} to,

\begin{equation}
ds^{2}=-\frac{y^{2}}{y^{2}+2m}dt^{2}+ 4\left(y^{2}+2m\right)dy^{2}
+\left(y^{2}+2m\right)^{2}(d\theta^{2}+\sin^{2}\theta d\phi^{2}).
\label{E-R}
\end{equation}
With $-\infty<y<\infty$, these coordinates omit the region inside the event horizon, $0<r<2m$, and twice cover the asymptotically flat region, $r\ge 2m$. The region near $y=0$ is the bridge connecting the two asymptotically flat regions close to $y=\infty$ and $y=-\infty$.  While the Einstein-Rosen bridge was not successful as a model for particles, it emerged as the prototype wormhole in gravitational physics, and eventually led to the study of traversable wormholes (see, e.g.  \cite{MT88, visser}, and for visual appearances, \cite{Muller1}).\\ 

\noindent It may be seen from Eq. \eqref{a1}, that if $r<2m$, then $r$ is a timelike coordinate, $t$ is spacelike, and within the event horizon, Schwarzschild spacetime is not static.  A test particle within the event horizon moves inexorably to smaller $r$ values because the flow of time is toward $r=0$.  Therefore, to understand the dynamics of the Einstein-Rosen bridge, or Schwarzschild wormhole, we require a spacetime that includes not only the two asymptotically flat regions associated with Eq. \eqref{E-R}, but also the region near $r=0$.  \\

\noindent For this purpose, the maximal extension of Schwarzschild spacetime, published in 1960 by Kruskal \cite{kruskal} and Szekeres, \cite{szekeres} along with their global coordinate system\footnote{ We refer to this coordinate system as ``Kruskal coordinates'' for simplicity.}, play an important role. The maximal extension includes not only the interior and exterior of the Schwarzschild black hole (covered by the coordinates of Eq. \eqref{a1}), but also a second copy of the exterior, as well as a region surrounding a white hole, from which particles may emerge but not enter (see Fig. \ref{CollasKleinFig02} below).\\  

\begin{figure}[!h]
  \begin{center}
    \includegraphics[height=1in, width=4.5in]{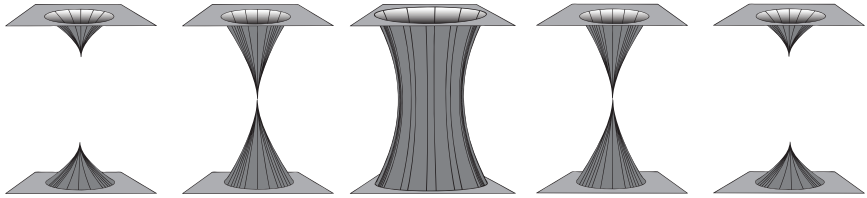}
     \end{center}
   \caption{Schematic for the dynamical evolution of the Schwarzschild wormhole.  From left to right: Kruskal time $v<-1$ prior to the formation of the bridge; Formation at $v=-1$; Maximum extent at $v=0$ (middle); Separation at $v=1$; Post separation at Kruskal time $v>1$ (far right).}
  \label{CollasKleinFig01}
\end{figure}

\noindent In 1962, Fuller and Wheeler \cite{FW62} employed Kruskal coordinates to describe the geometry of the Schwarzschild wormhole, and showed that it is nontraversable, even by a photon.  Their paper includes sketches of a sequence of wormhole profiles for particular spacelike slices, illustrating the formation, collapse, and subsequent ``pinching-off'' of the wormhole, but calculations for the embeddings in $\mathbb{R}^{3}$ were not provided. The nontraversability of the Schwarzschild wormhole makes the embeddings of its stages tricky, and this perhaps accounts for the dearth of explanation in the literature.  Indeed, while several widely used general relativity textbooks include qualitative descriptions and sketches of the dynamics of the Schwarzschild wormhole (or its profiles) similar to those in Fig. \ref{CollasKleinFig01}, we are not aware of any published explanation for the calculations of embeddings. \cite{fnote 1}\\

\noindent In this article, using elementary methods, we show how to embed spacelike slices of the Schwarzschild wormhole in $\mathbb{R}^{3}$.  The embeddings may be thought of as snapshots of the wormhole in its various stages, as measured by a party of explorers (represented by test particles) at fixed times, along particular spacelike slices of spacetime.  The most natural spacelike slices, from the point of view of Kruskal coordinates, are slices of constant Kruskal time $v$, and we consider those first.  However, other slices are also possible and give rise to interesting geometries, as we illustrate in the sequel. The calculations and examples, apart from supplying the missing material to interested physicists, are also suitable for general relativity and differential geometry courses.\\ 

\noindent This article is organized as follows. In Sec. \ref{Kruskal}, we review Kruskal coordinates for the maximal extension of Schwarzschild spacetime, display the metric in this coordinate system, and give a qualitative description of the dynamics of the Schwarzschild wormhole.  In Sec. \ref{Embed}, we develop a general method for embedding space slices as surfaces of revolution in $\mathbb{R}^{3}$.  Section \ref{Const. v} applies that method to embeddings of constant Kruskal times between $-1$ and $1$ in order to reveal the full dynamics of the wormhole from  formation to collapse.  At Kruskal times $v>1$, the two universes, connected by the wormhole at earlier times, $|v| < 1$, have separated into two connected components. In Sec. \ref{Sing. v} we show how to deal with a technical issue in order to embed space slices of nearly constant Kruskal time $v>1$ (or $v<-1$).    Then in Sec. \ref{Var. v} we show by example how more general embeddings can be carried out, including an embedding consisting of three separated components. Concluding remarks are given in Sec. \ref{Concl}.

\section[Kruskal]{Kruskal coordinates}\label{Kruskal}

\noindent Kruskal coordinates $u, v$ and Schwarzschild coordinates are related by, 
\begin{equation}
\label{K1}
u^{2}-v^{2}=\left(\frac{r}{2m}-1\right)e^{\frac{r}{2m}}\,,\;\;\;\;\;r\geq 0\,,
\end{equation}
and,
\begin{equation}
\label{K2}
\frac{2uv}{u^{2}+v^{2}}=\tanh\left(\frac{t}{2m}\right),
\end{equation}

\noindent where $t$ and $r$ are the time and radial Schwarzschild coordinates respectively of Eq. \eqref{a1}, and the angular coordinates are unchanged.  Coordinate $v$ is timelike and it follows from Eq. \eqref{K2} that constant ratios $v/u$ correspond to constant Schwarzschild $t$. In Kruskal coordinates, the Schwarzschild metric \eqref{a1} becomes,
\begin{equation}
\label{K3}
ds^{2}=\frac{32m^{3}}{r}\,e^{-\frac{r}{2m}}\left(-dv^{2}+du^{2}\right)+r^{2}\left(d\theta^{2}+\sin^{2}\theta d\phi^{2}\right),
\end{equation}

\noindent where $r=r(u,v)$ is determined implicitly by Eq. \eqref{K1}, and may be expressed in terms of the \textit{Lambert $W$ function} \cite{CK96} as follows.  Dividing both sides of Eq. \eqref{K1} by $e$ we may write
\begin{equation}
\label{E5}
\zeta=W(\zeta)e^{W(\zeta)}\,,
\end{equation}

\noindent where in the present case,
\begin{equation}
\label{E5a}
\zeta=\frac{u^{2}-v^{2}}{e}\,.
\end{equation}

\noindent The Lambert $W$ function, $W(\zeta)$, is defined to be the function satisfying Eq. \eqref{E5}, i.e., it is the inverse function to $\zeta(x)=xe^x$.  Then,
\begin{equation}
\label{E6}
W\left(\frac{u^{2}-v^{2}}{e}\right)=\frac{r(u,v)}{2m}-1\,,
\end{equation}

\noindent and
\begin{equation}
\label{E7}
r(u,v)=2m\left[1+W\left(\frac{u^{2}-v^{2}}{e}\right)\right]\,.
\end{equation}
 
\begin{figure}[!h]
  \begin{center}
    \includegraphics[width=2.5in]{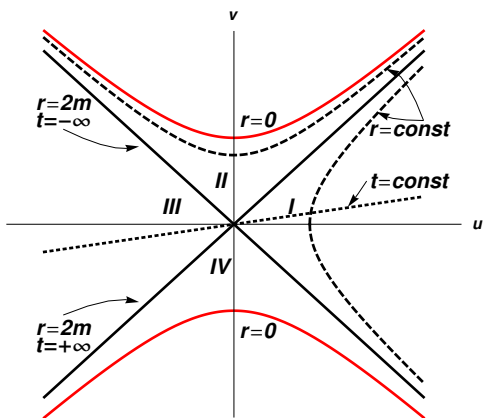}
     \end{center}
   \caption{Kruskal diagram of the Schwarzschild geometry.}
  \label{CollasKleinFig02}
\end{figure}

\noindent Fig. \ref{CollasKleinFig02} shows the maximally extended Schwarzschild spacetime in terms of Kruskal coordinates, with the angular coordinates suppressed, so that each point in the diagram represents a $2$-sphere.\\

\noindent The original Schwarzschild coordinates cover only regions I and II in Fig. \ref{CollasKleinFig02}.   Region II is the interior of the black hole.  Region III is a copy of region I, and region IV is the interior of a white hole, which lies in the past of any event.  The singularity at $r=0$ is given by the ``singularity hyperbola,'' $v_{s}(u)=\pm\sqrt{1+u^{2}}$. Lines making 45$^{\circ}$ angles with the coordinate axes are null paths. Additional details may be found, for example, in \cite{MTW73} and \cite{HEL06}.\\

\noindent The dynamics of the  wormhole may now be described qualitatively using Figures \ref{CollasKleinFig01} and \ref{CollasKleinFig02}.   Fig. \ref{CollasKleinFig02} shows two distinct, asymptotically flat Schwarzschild manifolds or ``universes,'' one consisting of regions I and II, and the other of regions III and IV.  Regions I and III are asymptotically flat, or Minkowskian far from the singularities. They may be identified as the top and bottom horizontal planes (respectively) in each diagram of Fig. \ref{CollasKleinFig01}.\\ 

\noindent When $v<-1$, the two universes are disconnected, each containing an infinite curvature singularity at $r=0$.   At time $v=-1$ the singularities join to form a nonsingular bridge.  As the universes evolve, the bridge increases, until $v=0$. At this instant, the wormhole has maximum width, with the observer at $u=0$ exactly on the event horizon, $r=2m$. As $v$ increases, the observer at $u=0$ enters the region $r<2m$, and the bridge narrows.  By the time $v=1$, the bridge pinches off, and the two separating universes just touch at $r=0$.   Thereafter the universes again separate, each with its singularity at $r=0$.\\ 

\noindent By examining paths of light rays, it may be seen from Fig. \ref{CollasKleinFig02} that no particle or photon can pass from region I to III or vice versa. Thus the wormhole is nontraversable.  However, before perishing, an observer who has fallen through the event horizon of the black hole, into Region II, could in principle see light signals from the other universe through the throat of the wormhole in this spacetime.\\

\noindent In the following section, we develop the mathematical machinery needed to carry out the embeddings of the various stages of the Einstein-Rosen bridge, described only qualitatively here.\\

\section[Embed]{Embedding wormholes as surfaces of revolution in $\mathbb{R}^{3}$}\label{Embed}

\noindent A particular choice of $v=v(u)$ as a function of $u$, with $\mid dv/du\mid<1$, selects a spacelike hypersurface (or a slice) through spacetime, and the projection of the metric of Eq. \eqref{K3} onto this hypersurface is,
\begin{equation}
\label{K4}
ds^{2}=\frac{32m^{3}}{r}\,e^{-\frac{r}{2m}}\left[1-\left(\frac{dv}{du}\right)^{2}\right]du^{2}+r^{2}\left(d\theta^{2}+\sin^{2}\theta d\phi^{2}\right)\,.
\end{equation}

\noindent Without loss of generality we shall consider slices with $v(u)\geq 0$.  For example, choosing $v(u)=\text{constant}$, for a range of constants, reveals  the dynamics of the wormhole depicted schematically in Fig. \ref{CollasKleinFig01}, but other choices, as we discuss in Sect. 6, are also possible.  As we explain below, slices with spacetime points with coordinate $v>1$, require special care near the singularity.\\

\noindent In order to produce embeddings of two dimensional space slices of the wormhole in $\mathbb{R}^{3}$, we make the restriction  $\theta=\pi/2$.  Specializing to the equatorial plane results in no loss of generality because of the spherical symmetry of Schwarzschild spacetime. With this constraint the metric of Eq. \eqref{K4} becomes,
\begin{equation}
\label{K5}
ds^{2}=\frac{32m^{3}}{r}\,e^{-\frac{r}{2m}}\left[1-\left(\frac{dv}{du}\right)^{2}\right]du^{2}+r^{2}d\phi^{2}\,.
\end{equation}

\noindent Our strategy is to identify Eq. \eqref{K5} as the metric of a surface of revolution in $\mathbb{R}^{3}$ induced by the Euclidean metric.  To that end, consider the profile (or generating) curve, $\alpha(u)=\left(f(u),0,z(u)\right)$, and the associated chart for a surface of revolution,
\begin{equation}
\label{E1}
\mathbf{x}(u,\phi)=\left(f(u)\cos\phi,\,f(u)\sin\phi,\,z(u)\right)\,.
\end{equation}

\noindent The induced metric \cite{O'N06} on the surface of revolution is, 
\begin{equation}
\label{E2}
ds^{2}=\left[\left(f^{\prime}(u)\right)^{2}+\left(z^{\prime}(u)\right)^{2}\right]du^{2}+f^{2}(u)d\phi^{2}\,.
\end{equation}

\noindent Comparing Eqs. \eqref{K5} and \eqref{E2}, we make the identifications,
\begin{equation}
\label{E3}
f(u)\equiv f(u,v(u))=r(u,v(u))\,,
\end{equation}

\noindent and
\begin{equation}
\label{E4}
\left(f^{\prime}(u)\right)^{2}+\left(z^{\prime}(u)\right)^{2}=\frac{32m^{3}}{r}\,e^{-\frac{r}{2m}}\left[1-\left(\frac{dv}{du}\right)^{2}\right]\,.
\end{equation}

\noindent To find $z(u)$, we begin by differentiating both sides of Eq. \eqref{K1} to get,
\begin{equation}
\label{R1}
2\left(u-v\frac{dv}{du}\right)du=\left(\frac{r}{4m^{2}}\right)e^{\frac{r}{2m}}dr\,,
\end{equation}

\noindent or
\begin{equation}
\label{R2}
du=\frac{\left(\frac{r}{8m^{2}}\right)e^{\frac{r}{2m}}}{\left(u-v\frac{dv}{du}\right)}\,dr\,.
\end{equation}

\noindent Now, assuming that $v(u)$ is chosen in such a way that $u$ can be expressed as a function of $r$ (through Eq.  \eqref{K1}), we may rewrite the metric from Eq. \eqref{K5} as,
\begin{equation}
\label{R3}
ds^{2}=\frac{32m^{3}}{r}\,e^{-\frac{r}{2m}}\left[1-\left(\frac{dv}{du}\right)^{2}\right]\left[\frac{\left(\frac{r}{8m^{2}}\right)e^{\frac{r}{2m}}}{\left(u-v\frac{dv}{du}\right)}\right]^{2}dr^{2}+r^{2}d\phi^{2}\,.
\end{equation}

\noindent The coefficient, $g_{rr}$, of $dr^{2}$, simplifies to
\begin{equation}
\label{R4}
g_{rr}=\left(\frac{r}{2m}\right)e^{\frac{r}{2m}}\frac{\left[1-\left(\frac{dv}{du}\right)^{2}\right]}{\left[u-v\frac{dv}{du}\right]^{2}}\,.
\end{equation}

\noindent From Eq. \eqref{E4}, and choosing the positive square root, we find that
\begin{equation}
\label{E9}
\frac{dz}{du}=\left[\frac{32m^{3}}{r}e^{-\frac{r}{2m}}\left[1-\left(\frac{dv}{du}\right)^{2}\right]-\left(f^{\prime}(u)\right)^{2}\right]^{\frac{1}{2}}\,,
\end{equation}

\noindent and from Eqs. \eqref{E3}, and \eqref{R2} we have that
\begin{equation}
\label{E10}
\frac{df}{du}=\frac{dr}{du}=\frac{8m^{2}\left(u-v\frac{dv}{du}\right)e^{-\frac{r}{2m}}}{r}\,.
\end{equation}

\noindent So Eq. \eqref{E9} takes the form,
\begin{equation}
\label{E11}
\frac{dz}{du}=\left[\frac{32m^{3}}{r}e^{-\frac{r}{2m}}\left[1-\left(\frac{dv}{du}\right)^{2}\right]-\left[\frac{8m^{2}\left(u-v\frac{dv}{du}\right)e^{-\frac{r}{2m}}}{r}\right]^{2}\right]^{\frac{1}{2}}\,.
\end{equation}
\vspace{0.3cm}

\noindent Eqs. \eqref{E4}, \eqref{E10}, and \eqref{E11}, then also give a useful  expression for $dz/dr$,
\begin{equation}
\label{E12}
\left(\frac{z^{\prime} (u)}{f^{\prime} (u)}\right)^{2}=\left(\frac{dz}{dr}\right)^{2}=g_{rr}-1\,.
\end{equation}

\noindent Thus, we require $g_{rr}\geq 1$ (see \cite{LPPT75} for a similar observation).

\section[Const. v]{Constant Kruskal time embeddings before separation of the universes}\label{Const. v}

\noindent In this section we consider space slices of the form $v(u)=v_{0}$, for any  constant $v_{0}\in [-1,1]$.  This interval of Kruskal coordinate times captures the full evolution of the Schwarzschild wormhole, and the slices constructed in this section are exact versions of the schematic diagrams in Fig. \ref{CollasKleinFig01}.\\

\noindent For ease of exposition, here and below we will take $m=1$.  The embeddings may be carried out using the machinery developed in the previous section. From Eqs. \eqref{E11} and \eqref{E12},
\begin{align}
z(u)&=\int\limits_{0}^{\;u}\frac{dz}{du}du=\int\limits_{r(0)}^{\;r(u)}\frac{dz}{du}\frac{du}{dr}\,dr\,,\label{E12.1} \\
&=\int\limits_{r(0)}^{\;r(u)}\sqrt{g_{rr}-1}\,dr\,,\label{E12.2}\\
&=\int\limits_{r(0)}^{\;r(u)}\left[\frac{2\left(e^{\frac{r}{2}}-v_{0}^{2}\right)}{re^{\frac{r}{2}}-2\left(e^{\frac{r}{2}}-v_{0}^{2}\right)}\right]^{\frac{1}{2}}dr\,,\label{E12.3}
\end{align}

\begin{figure}[!h]
  \begin{center}
    \includegraphics[height=2.5in, width=2.0in]{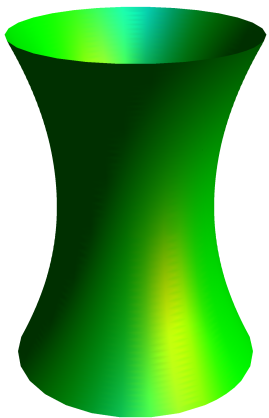}
     \end{center}
   \caption{The wormhole embedding for Kruskal time $v=v_{0}=0.75$.}
  \label{CollasKleinFig03}
\end{figure}

\begin{figure}[!h]
  \begin{center}
    \includegraphics[height=2.5in, width=2.0in]{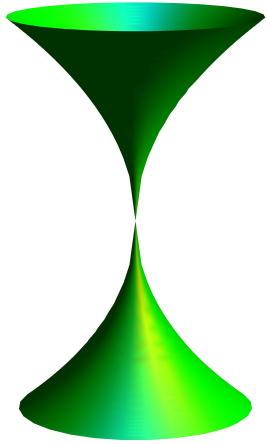}
     \end{center}
    \caption{The wormhole at the instant of pinching, i.e., $v=v_{0}=1$.}
  \label{CollasKleinFig04}
\end{figure}

\noindent where $r(0)$ is given by Eq. \eqref{E7}, i.e.,
\begin{equation}
\label{E12.5}
r(0)=r(0,v_{0})=2\left[1+W\left(\frac{-v_{0}^{2}}{e}\right)\right]\,.
\end{equation}

\noindent The integral of Eq. \eqref{E12.3} can be calculated in closed form for the slice $v_{0}=0$.  This is the original Einstein-Rosen bridge, and it is typically the only example treated in detail in textbooks. \cite{fnote 1} The other embeddings require numerical integrations, and for that purpose  we find it convenient to use the first integral in Eq. \eqref{E12.1}.  For the limiting cases $v_{0}=\pm 1$, the slice meets the singularity tangentially at $u=0$, and that results in an improper  integral for $z(u)$, which, however, as we show in the next section, is convergent.  Fig. \ref{CollasKleinFig03} shows the embedding of the wormhole at Kruskal time $v_{0}=0.75$, and Fig. \ref{CollasKleinFig04} shows it at $v_{0}=1$, the instant of pinching off.  

\section[Sing. v]{Separated universes at (nearly) constant Kruskal times}\label{Sing. v}

\noindent In this section we consider the embedding of the wormhole for $v(u)=v_0$ for constant Kruskal time $v_0$, but with $v_0^2>1$.  When $v_0<-1$, the wormhole has not yet formed, and for $v_0>1$, the wormhole has already formed, then  pinched off, and the two universes have separated.  We will see that the restriction that $v(u)$ is constant must be relaxed at spacetime points close to the singularity if the embeddings are to include such points.\\

\noindent Since the singularity at $r=0$ in Kruskal coordinates is given by the singularity hyperbola $u^{2}-v^{2}=-1$, our function $v(u)=v_0$ is defined only for the values of $u$ such that the pair $(u, v_0)$ (with the angular coordinates suppressed) lies within the Kruskal coordinate chart, i.e, for values of $u$ that satisfy $u^2 > v_0^2 -1$.\\

\noindent However, the allowable values of $u$ must be further restricted in order to satisfy the inequality $g_{rr}\geq 1$ that follows from Eq. \eqref{E12}.  To see this, observe that when $v(u)=v_0$ Eq. \eqref{R4} reduces to,
\begin{equation}
\label{R4'}
g_{rr}=\frac{re^{\frac{r}{2}}}{2u^2}.
\end{equation}

\noindent Combining Eq. \eqref{R4'} with Eqs. \eqref{K1} and \eqref{E12} we obtain,
\begin{equation}
\label{E12'}
\left(\frac{dz}{dr}\right)^{2}=\frac{2\left(e^{\frac{r}{2}}-v_{0}^{2}\right)}{re^{\frac{r}{2}}-2\left(e^{\frac{r}{2}}-v_{0}^{2}\right)}\,.
\end{equation}

\noindent From this last equation we see that $r$ must be restricted to those values for which the right-hand side of Eq. \eqref{E12'} is non-negative.  It is easily checked that the denominator, $D(r)$, of the right-hand side of Eq. \eqref{E12'} satisfies $D(0)>0$ and $D'(r)>0$ for $r>0$. Therefore, $D(r)>0$ for all $r>0$, and the sign of the right-hand side of Eq. \eqref{E12'} is determined solely by the numerator.  Therefore we must have $r>2\ln{v_{0}^2}$.  Then, using Eq. \eqref{K1} again, we find that, 
\begin{equation}
\label{C3'}
u^2\geq v_{0}^2 \ln{v_{0}^2}\equiv u_0^2.
\end{equation}

\noindent That is, the embedding for $v(u)=v_0$ is possible for $u^2\geq u_0^2=v_{0}^2 \ln{v_{0}^2}$ only \cite{fnote 3}. In other words, an embedding is not possible for the space slice given by the horizontal line, $v(u)=v_{0}>1$, that extends all the way to the singularity hyperbola, $r=0$. Surprisingly, the horizontal line must be bounded away from $r=0$ for the embedding to be possible.\\

\begin{figure}[!h]
  \begin{center}
    \includegraphics[width=2.5in]{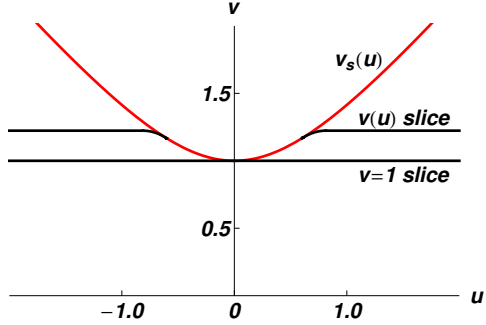}
     \end{center}
   \caption{Typical intersections of spacelike slices with the singularity hyperbola $v_{s}(u)$. Horizontal lines with $v>1$ cannot reach the singularity hyperbola, $r=0$; they must be extended inward beyond $u_{0}$ by lines or curves tangent to the hyperbola.}
  \label{CollasKleinFig05}
\end{figure}

\begin{figure}[!h]
  \begin{center}
    \includegraphics[width=2.0in]{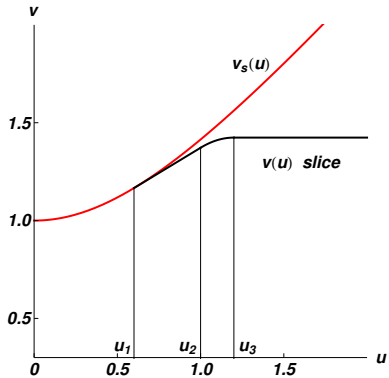}
     \end{center}
   \caption{A smooth, nearly constant space slice $v(u)$ (for $u>0$) tangential to the singularity.}
  \label{CollasKleinFig06}
\end{figure}

\noindent Nevertheless, our embedding can be extended to include smaller values of $|u|$, thus capturing more of the Schwarzschild spacetime, if we weaken the restriction that $v(u)$ is a constant function. To that end, we shall extend  $v(u)$ so that it is a linear function of $|u|$, with positive slope, for values of $u$ with $u^2 < u_0^2$.  To avoid restrictions of domain analogous to Eq. \eqref{C3'}, we consider linear functions that are tangent to the singularity hyperbola, $v_{s}(u)$, as illustrated in Fig. \ref{CollasKleinFig05} (see also Fig. \ref{CollasKleinFig06}).  We proceed now to show that this choice enables us to include spacetime points arbitrarily close to the singularity.  For simplicity, we restrict our attention to the portion of the spacetime with $u\geq0$.  The extension to negative values of $u$ is straightforward as we explain below Eq. \eqref{D2}.\\

\noindent Let $v_{1}(u)=a_{1}u+b_{1}$ be a linear function that meets the singularity hyperbola, $v_{s}(u)=\sqrt{1+u^{2}}$, tangentially at $u=u_{1}\geq{0}$.  Then a simple calculation shows that,
\begin{equation}
\label{P2}
a_{1}=\frac{u_{1}}{\sqrt{1+u_{1}^{2}}}\,,\;\;\;\;\;b_{1}=\frac{1}{\sqrt{1+u_{1}^{2}}}\,.
\end{equation}

\noindent With the domain of $v_{1}(u)$ temporarily taken to be $[u_{1}, \infty)$, we choose the antiderivative, $z(u)$ of Eq. \eqref{E11}, as,
\begin{equation}
\label{D1}
z(u)=\int\limits_{u_{1}}^{\;u}\frac{dz}{du}du=\int\limits_{0}^{\;r(u)}\frac{dz}{dr}\,dr=\int\limits_{0}^{\;r(u)}\sqrt{g_{rr}-1}\,dr\,,
\end{equation}

\noindent where, $(u_{1},v_{1})$, is the point of tangency.  From Eq. \eqref{R4}, with $v=v_{1}(u)$, we have that,
\begin{equation}
\label{P4}
g_{rr}=\frac{r}{2}\,e^{\frac{r}{2}}\,\frac{1+u_{1}^{2}}{(u-u_{1})^{2}}\,.
\end{equation}

\noindent It is easily verified that,
\begin{align}
\frac{(u-u_{1})^{2}}{1+u_{1}^{2}}&=u^{2}-v_{1}^{2}(u)+1\label{P5a} \\
&=1+e^{\frac{r}{2}}\left(\frac{r}{2}-1\right),\label{P5b}
\end{align}

\noindent where in the last step we made use of Eq. \eqref{K1}.  Substituting Eq. \eqref{P5b} in Eq. \eqref{P4}, we may write the last integral in Eq. \eqref{D1} as,
\begin{equation}
\label{P5}
z(r(u))=\int\limits_{0}^{\;r(u)}\left[\frac{2\left(e^{\frac{r}{2}}-1\right)}{r\,e^{\frac{r}{2}}-2\left(e^{\frac{r}{2}}-1\right)}\right]^{\frac{1}{2}}\,dr\,.
\end{equation}

\begin{remark} The integral of Eq. \eqref{P5} is the same as that in Eq. \eqref{E12.3}, when $v(u)\equiv v_0=1$, for which case $v(u)$ is tangent to the singularity hyperbola at $r=0$.
\end{remark}

\noindent To see that the integral in Eq. \eqref{P5} converges and is real, we first note that the same argument applied to Eq. \eqref{E12'} shows that  
\begin{equation}
\label{E12''}
\left(\frac{dz}{dr}\right)^{2}\equiv\frac{2\left(e^{\frac{r}{2}}-1\right)}{re^{\frac{r}{2}}-2\left(e^{\frac{r}{2}}-1\right)}>0\,,
\end{equation}

\noindent when $r>0$.  Then, a calculation using L'H\^opital's rule shows that,
\begin{equation}\label{limit}
\lim_{\,r\rightarrow0^+}\left[r\times\left(\frac{dz}{dr}\right)^{2}\right]=4\,,
\end{equation}

\noindent from which it follows that,
\begin{equation}
\label{compare}
\frac{dz}{dr}=\left[\frac{2\left(e^{\frac{r}{2}}-1\right)}{r\,e^{\frac{r}{2}}-2\left(e^{\frac{r}{2}}-1\right)}\right]^{\frac{1}{2}}<\frac{C}{\sqrt{r}}\,,
\end{equation}

\noindent for $0<r<\delta$, where $C$ and $\delta$ are some positive constants.  Since the right-hand side of Eq. \eqref{compare} is integrable, it follows that the integral in Eq. \eqref{P5} converges. \\

\noindent We may now extend our original constant function $v(u)=v_0$ by redefining $v(u)$ to be $v_1(u)$ for $u_1 < u \leq u_0$, and $v(u)=v_0$ for $u>u_0$, for which an embedding function $z(u)$ may be well defined. However, this extended space slice, $v(u)$, fails to be differentiable at $u_0$, where it has a kink.  To remedy this, we may construct spacelike slices consisting of a straight line segment, $v=v_{1}(u)$, tangent to the singularity hyperbola, $v_{s}(u)$, at $(u_{1},v_{1})$ and joined, with the desired degree of smoothness, to another curve, $v_{2}(u)$, (satisfying the spacelike condition, $|dv_{2}/du|<1$), which in turn is joined to a constant function, $v_{3}(u)$.  The typical situation we have in mind is illustrated in Fig. \ref{CollasKleinFig06}, where the straight line segment, $v_{1}(u)$, is joined to a parabola, $v_{2}(u)$, which in turn is joined at its maximum to a horizontal straight line, $v_{3}(u)\equiv v_0$.  In the composite slice the joints are of class $C^{1}$, i.e., continuously differentiable.\\

\noindent For this composite space slice, we have a well defined function $z(u)$ given by Eq. \eqref{D1} for $u\geq 0$, where as before, $(u_{1},v_{1})$ is the point where $v(u)$ meets the singularity hyperbola.  For $u\leq 0$,  $z(u)$ is defined by
\begin{equation}
\label{D2}
z(u)=\int\limits_{-u_{1}}^{\;u}\frac{dz}{du}du\,,\;\;\;u\leq -u_{1}\leq 0\,.
\end{equation}

\noindent Note that since $v(u)$ is an even function of $u$, it follows that $r(u,v(u))$ and $dz(u)/du$ are also even functions, and consequently $z(-u)=-z(u)$, i.e., $z(u)$ is odd.  This observation simplifies the numerical calculations.\\

\begin{figure}[!h]
  \begin{center}
    \includegraphics[width=2.0in]{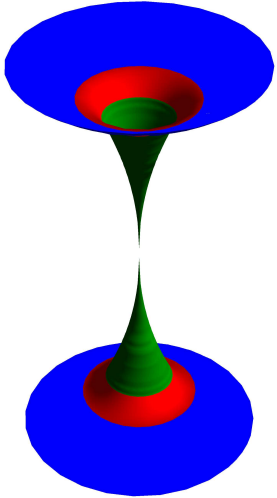}
      \end{center}
     \caption{The collapsed wormhole corresponding to Fig. \ref{CollasKleinFig06}.  The distance between the pinches is arbitrary.}
   \label{CollasKleinFig07}
\end{figure}

\noindent Fig. \ref{CollasKleinFig07} shows an embedding for a space slice, $v(u)$, consisting of the three parts illustrated in Fig. \ref{CollasKleinFig06}. The straight line, $v_{1}(u)$ of Eq. \eqref{P2}, is tangent to the singularity hyperbola at $u_{1}=0.6$.  The parabola, $v_{2}(u)=v_{1}(u)$, at $u=u_{2}=1.0$ and their slopes match there.  The parabola, $v_{2}(u)$, is joined at its maximum, $u_{3}=1.2$, to the horizontal straight line, $v_{3}(u)$, whose constant value is approximately $1.424$ (and thus the slopes match there also). We have ensured that the joints are of class $C^{1}$.\\

\begin{figure}[!h]
  \begin{center}
    \includegraphics[width=2.0in]{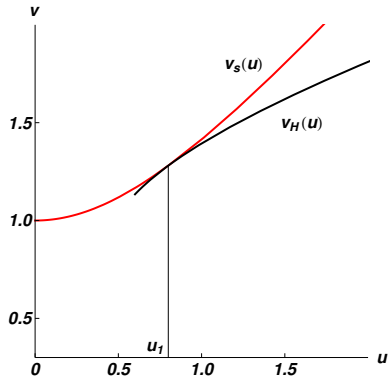}
      \end{center}
     \caption{The hyperbola $v_{H}(u)$ slice for $u>0$.}
   \label{CollasKleinFig08}
\end{figure}

\begin{figure}[!h]
  \begin{center}
    \includegraphics[width=2.0in]{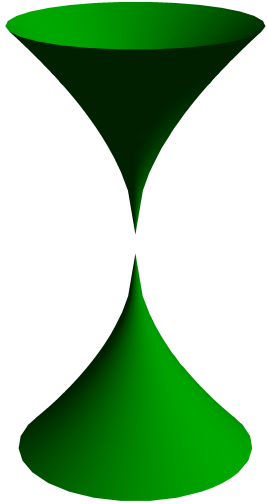}
     \end{center}
   \caption{The wormhole of the slice $v_{H}(u)$.  The distance between the pinches is arbitrary.}
  \label{CollasKleinFig09}
\end{figure}

\noindent Another space slice serves both as an alternative to the previous example and as a transition to the next section. Let,
\begin{equation}
\label{H1}
v_{H}(u)=k+\frac{b}{a}\sqrt{\left(u-h\right)^{2}-a^{2}}\,,
\end{equation}

\noindent where we choose $a=4$, $b=1$.  The values $h\approx -3.5648$, and $k\approx 0.84394$ are then determined by our additional requirement that $v_{H}(u)$ is tangent to the singularity hyperbola at $u_{1}=0.8$, as shown in Fig.  \ref{CollasKleinFig08}. This space slice is infinitely differentiable, and analogous to the derivations of Eqs \eqref{limit} and \eqref{compare} for the straight line tangent, it can be shown for the present example that,
\begin{equation}
\label{C2}
\frac{dz}{du}<\frac{C}{\sqrt{u-u_{1}}}\,,
\end{equation}

\noindent for $0 < u-u_{1} < \delta$, where $C$ and $\delta$ are some positive constants. \cite{fnote 4}  Therefore the integral in Eq. \eqref{D1} converges. The corresponding embedding of the collapsed wormhole is shown in Fig. \ref{CollasKleinFig09}.

\pagebreak 

\section[Var. v]{Three component embeddings}\label{Var. v}

\noindent  In the previous sections our focus was on embeddings of the Schwarzschild wormhole for constant, or nearly constant, Kruskal times.  However, as illustrated in the last example of the previous section, smooth embeddings are also readily available for an unlimited choice of other non constant space slices.  As discussed in the introduction, a party of explorers (of negligible mass) is free to choose any space slice, $v(u)$, including one that pushes one region of space faster ahead in Kruskal time than other regions.\\

\noindent As an illustration, consider a family of space slices defined for all values of $u$, whose graphs are again hyperbolae.  Let,
\begin{equation}
\label{F}
v(u, v_{0})=v_{0}+\frac{p}{q}\sqrt{\left(u^{2}+q^{2}\right)}\;.
\end{equation}
where  $p$ and $q$ are positive constants, and $v_{0}$ is a parameter whose variation may be regarded as representing the time evolution of this family of space slices.  For concreteness, we take $p=q/2$ and $q=0.1$.\\

\begin{figure}[!h]
  \begin{center}
    \includegraphics[width=2.5in]{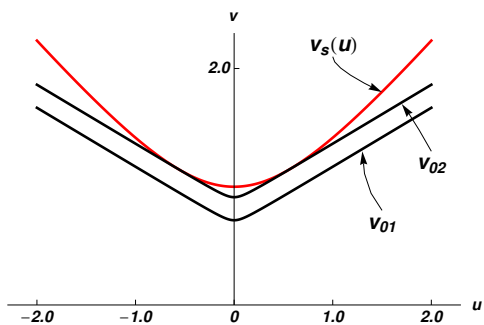}
     \end{center}
   \caption{The hyperbolae $v(u,v_{0})$ slices.}
  \label{CollasKleinFig10}
\end{figure}

\begin{figure}[!h]
  \begin{center}
    \includegraphics[width=2.0in]{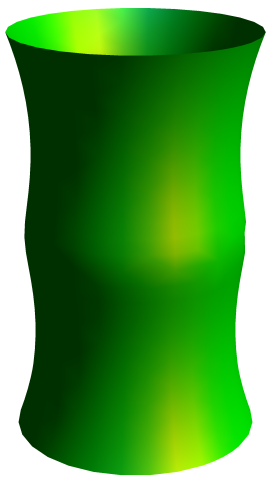}
     \end{center}
   \caption{The wormhole of the slice $v(u,v_{0})$, with $v_{0}=v_{01}$.}
  \label{CollasKleinFig11}
\end{figure}

\noindent We may again use Eq. \eqref{E11}  to calculate $z(u)$.  The integrations present no special difficulties, but must be done numerically. For the case that $v_{0}\equiv v_{01}=0.6$, the slice and corresponding wormhole are shown in Figs. \ref{CollasKleinFig10} and \ref{CollasKleinFig11}.\\

\begin{figure}[!h]
  \begin{center}
    \includegraphics[height=3.0in, width=2.0in]{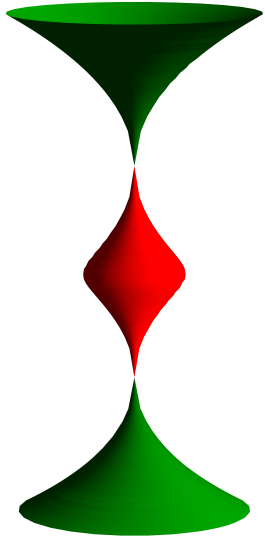}
     \end{center}
   \caption{The wormhole of the slice $v(u,v_{0})$, with $v_{0}=v_{02}$.}
  \label{CollasKleinFig12}
\end{figure}

\noindent Now let $v_{0}$ increase until $v(u,v_{0})$ has two points of tangency with the singularity hyperbola, as shown in Fig. \ref{CollasKleinFig10}.  This occurs when $v_{0}\equiv v_{02}=3\sqrt{33}/20$. As $v_{0}$ increases to this value, the cylinder gradually pinches off at two values of $v$.  The corresponding embedding displayed in Fig.  \ref{CollasKleinFig12} shows a separation of universes into three components.  A qualitatively similar model of a collapsing dust star is discussed in detail in Box 32.1 of Ref. \cite{MTW73} (see figure on p. 856).

\pagebreak

\section[Concl]{Concluding remarks}\label{Concl}

\noindent Using Kruskal coordinates for the maximal extension of the Schwarzschild geometry, we have shown how to embed projections of the Schwarzschild wormhole as surfaces of revolution in $\mathbb{R}^{3}$ for a general class of spacelike slices $v(u)$.  A foliation of spacetime by a collection of surfaces $\{v(u)\}$ constitutes a definition of simultaneity.  A party of explorers of negligible mass, acting as a collection of test particles, is free to make any such choice, subject only to the constraint, $|dv/du|<1$. For slices of constant  Kruskal time, the embeddings trace the time evolution of the wormhole within the Kruskal coordinate system, providing snap shots from its formation at $v=-1$ to the pinching off at $v=1$, and beyond in the sense of Sec. \ref{Sing. v}.\\  

\noindent However, other choices of $v(u)$ can push one region of space faster ahead in coordinate time (i.e., Kruskal time) than others, at the option of the party of explorers.  In Sec. \ref{Var. v}, we showed how one choice of evolving hypersurfaces of simultaneity leads to a separation into three disjoint components, in the limit as the hypersurfaces approach the singularity tangentially.   More generally, multiple components arise through evolving hypersurfaces approaching $r=0$ at multiple tangent points.  Spacetime is four dimensional and the slices are only three dimensional, so the selection of space slices is unlimited.  There are no physically correct or incorrect choices.   What would an individual observer actually see in the vicinity of a Schwarzschild black hole?  The answer depends on the observer and direction of sight, but a discussion for an observer in circular motion around the black hole was given in \cite{Muller2}.\\

\noindent Other embedding strategies can also provide insights into the Schwarzschild geometry as well as the geometries of other spacetimes.  Embeddings of two-dimensional submanifolds in 3-dimensional Minkowski spacetime, $\mathbb{R}^{3}_{1}$, with a time coordinate serving as one of the two parameters,  were investigated in \cite{M99}, \cite{GMG04}, \cite{GH04}.  Each embedding of the Schwarzschild wormhole, whether in $\mathbb{R}^{3}$ or $\mathbb{R}^{3}_{1}$,  reveals only part of its intricate geometry. \\

\noindent \textbf{Acknowledgement.}  The authors would like to thank Louis Solis, of the Department of Art at California State University, Northridge, for Figure 1.\\


\begin{thebibliography}{99}
\bibitem{fnote 4} More generally one can show, using Eq. (4.22) of Ref. \cite{CK96}, that if a spacelike slice $v(u)$, has the Taylor expansion about a point $u_{1}$ given by, $v(u)=v_{s}(u_{1})+\left[u_{1}/v_{s}(u_{1})\right](u-u_{1})+\left[v^{\prime\prime}(u_{1})/2\right](u-u_{1})^{2}+O\left((u-u_{1})^{3}\right)$, and thus meets the singularity hyperbola, $v_{s}(u)$, tangentially at $u_{1}$, then $dz/du\sim C/\sqrt{|u-u_{1}|}$, in an open interval containing $u_{1}$, where $C$ is a positive constant, provided $v^{\prime\prime}(u_{1})<1/\left(1+u_{1}^{2}\right)^{3/2}$.
\end{thebibliography}
\end{document}